# Analytic model for sheath-plasma resonance in inverted fireballs

Subham Dutta[1], Johannes Gruenwald[2], Pralay Kumar Karmakar[1]
[1]Department of Physics, Tezpur University, Napaam, Tezpur-784028, Assam, India
[2]Gruenwald Laboratories GmbH, Taxberg 50, 5660 Taxenbach, Austria
Corresponding author E-mail: pkk@tezu.ernet.in, pkk170722@gmail.com

The sheath plasma resonance (SPR) in an inverted fireball (IFB) system is semi-analytically investigated in the framework of a generalized hydrodynamic isothermal model formalism. It incorporates the constitutive ionic fluid viscosity, inter-species collisions, and geometric curvature effects. The SPR stability is studied for an anodic (hollow, meshed) IFB for the first time against the traditional cathode-plasma arrangements of regular electrode (solid, smooth) fireballs. The SPR develops in the vicinity of a spherical electrode enclosed by a plasma sheath amid a given electric potential. A generalized linear quartic dispersion relation (DR) with diverse plasma multi-parametric coefficients is methodically derived using a standard normal mode analysis. The mathematical construct of the obtained DR roots confirms that there exists only one feasible nonzero frequency mode (emerging in the IFB). This root existence is confirmed both analytically and numerically. This consequent SPR creates trapped acoustic fluctuations in the IFB plasmas because of the internal reflections at the sheath plasma boundary. Also, sensible parametric changes in the SPR features, with both plasma density and viscosity, are seen. A local condition ($\omega \leq \omega_{spe}$) for the SPR excitation and its subsequent transition to collective standing wave-like patterns in the IFBs is illustrated. A fair corroboration of our results with the earlier SPR experimental observations of standing wave-like eigenmode patterns (evanescent) strengthens the reliability of our study alongside new applicability.

Subject Areas: Inverted fireball, Sheath plasma resonance, Sheath plasma instability

## I. INTRODUCTION

It is seen that a plasma fireball (FB) is a discharge phenomenon occurring both in laboratory and astrophysical-space plasma environments [1]. The two seemingly non-identical FBs (in terms of dimension and brightness) within their different environments (laboratory and astrophysical) and dissimilar geometrical structures manifest almost identical excitation dynamics. Their light emission occurs through active relaxation processes of the excited neutrals [1,2]. The difference, however, lies in the force causing the excitation of the neutrals between these two classes of fireballs. The laboratory plasma FBs develop due to electric field-induced inelastic electron-neutral collisions. Whereas the astronomic FBs form due to comets heating up while moving through Earth's atmosphere. The heat generated due to friction turns the vapors into plasma and the excitation of neutral gas molecules beneath the meteors release the energy in the form of visible light [2,3].

In principle, any electrode with adequately raised electrostatic potential (relative to the ionization and plasma potentials) in a surrounding plasma can create a plasma FB on its surface or within it, if a solid or hollow meshed anode is used, respectively [4,5]. However, in the latter case, it must be ensured that the grid constant is maximum twice the Debye length [6]. These two distinct types of laboratory fireballs based on their different anode morphology are designated as regular fireball (RFB) and inverted fireball (IFB), respectively. Due to the greater electrostatic potential, drifting electrons acquire velocities fast enough to excite or even ionize the neutrals through inelastic collisions. Thus, they form the plasma FB glow inside the sheath region around the anode [2]. The IFB is confined within a hollow grid anode (unlike solid anodes in RFB) [5]. Some secondary electrons and ions are also produced due to the additional ionization processes inside the IFB [7,8].

The spatial dimension of the laboratory plasma FBs is determined by the electron-ion flux balance (floating condition) through the plasma boundary [4]. It is noteworthy to add here that a highly nonlinear double layer (DL) forms around a FB. This DL gathers charges from the ambient plasma and accelerates them subsequently. It also detaches the sheath region from the ambient plasma [2]. It is pertinent to note that any imbalance of the particle flux across the DL may instigate various instabilities, which are, in fact, of great interest for both fundamental as well as applied research [9].

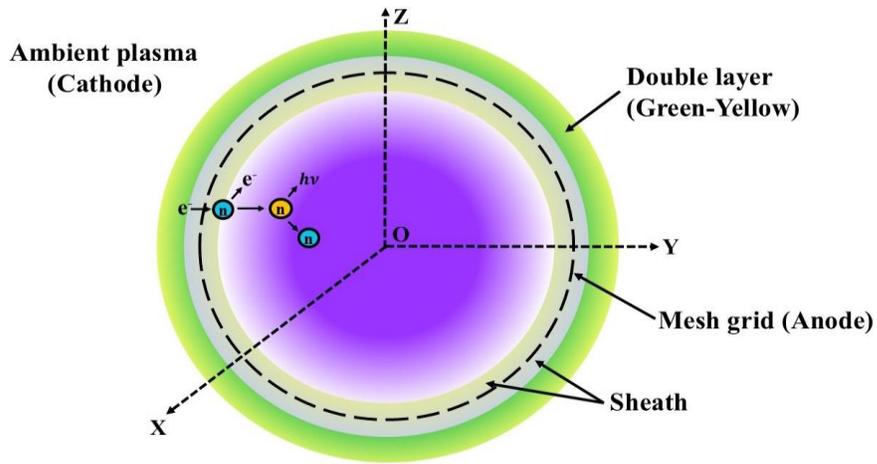

Fig. 1. A schematic of an IFB with spherical geometry. Electrons collide with neutrals within the IFB resulting in their excitation and ionization. Light and secondary electrons are emitted thereof.

Plasma FBs are first described in research conducted on plasma electrode interactions by Otto Lehmann [10] and Irving Langmuir et al. [11] in the early 20$^{th}$ century. However, more recent research on plasma FBs has been initiated by Stenzel et al. in 2008 [4,12]. The previous studies have dominantly reported the current-voltage relationships and geometrical variations of FBs [4]. The studies of FB instabilities gradually developed after 2010 [7-9,12-15]. The astrophysical relevance of plasma FBs also lies in associated instabilities. The knowledge of such instabilities can also be utilized in structure formations in astroplasmas under different conditions.

It may be noted that an electron-depleted sheath and a field-free cold plasma in an IFB system behave as a capacitor and inductor, respectively. A small-scale perturbation in the system can lead to unstable oscillations in the peripheral sheath and in the plasma region within the field free IFB. The sheath plasma resonance (SPR) can yield plasma oscillations resonating with the sheath [16,17]. The SPR in a system can be experimentally detected with a small amplitude microwave signal incident on the oscillating sheath and observing its reflected amplitude using directional couplers [17,18]. This resonance occurs when the electron transit time across the sheath is of the order of the inverse of electron plasma frequency ($\omega_{pe}$) and hence, determines the SPR frequency [16].

Apart from the SPR, there are some other instabilities excitable in a plasma-electrode system, such as, the secondary electron emission instability, the Rayleigh-Taylor instability, the two-stream instability, and so forth [2,19]. Most of these instabilities observed both in the laboratory as well as in astrophysics indicate the significance of research regarding instabilities and their linear and nonlinear behavior. In this work, we emphasize the SPR-induced oscillations in an IFB system under laboratory conditions. The SPR is also linked to the sheath plasma instability (SPI) discussed herein and in reference [17]. This work uses a linear perturbation formalism of the governing equations of an IFB system that finally leads to a quartic equation for the angular instability frequency $\omega$.

It may also be added that by increasing the dc bias of the IFB grid anode, the SPI creates numerous harmonics along with line broadening, hard thresholds, downward frequency shifting, and so forth. The various harmonics of the resonant oscillations are radiated as EM waves in the MHz-GHz frequency range from the anode, which also acts as a plasma antenna [17].

It is noteworthy that the SPR instabilities have so far been reported only in connection with classical RFBs [18]. Studies in IFB systems haven't been conducted yet to the best of our knowledge. Thus, the main motivation of the current work is devoted to studying such resonance instability phenomena within the IFBs. Besides, a bifluidic ansatz is used in this work for the first time to describe IFBs with variations in density, viscosity, and linear small-scale perturbation of the involved plasma parameters. This theoretical model recreates a few of the experimental outcomes on the IFBs reported previously in the literature.

## II. MODEL EQUATIONS

The formation of an IFB within a spherical mesh anode in a bifluidic (electronic-ionic) plasma is contemplated. These bifluidic plasma parameters are assumed to be consistent with former laboratory IFB experiments [1,2]. It must be noted that both RFB and IFB geometries are reported to vary from spherical to non-spherical shapes. Besides, fireball structures formed around a spherical anode can also deviate from their spherical geometry due applied magnetic fields [20-22].

We consider a simpler, magnetic field free fireball system (of spherical geometry) as shown in Fig. 1. The considered spherical symmetry of the IFB reduces the analysis into a one-dimensional (radial) problem. The individual dynamics of the two fluids (electrons-ions) fulfil the plasma conditions $((r,t) \gg (\lambda_D, \omega_p^{-1}))$. The local imbalance of charges within the double layer (DL) surrounding the IFB induces a non-neutral local charge number density ($n_e \neq n_i$). The constitutive number density fields in the electrical Poisson equation develop the corresponding potential dispersal as a useful closure property. The DR is finally obtained by decoupling the continuity and momentum equations amid the Poisson equation. The DR is solved both analytically (with a few reasonable approximations) and numerically (without any approximation). It turns out that in many cases the analytical approximation is sufficient to describe the IFB system very well. The unique analytical solution of the DR describes the angular frequency (and its harmonics) generated in the system. In the numerical solution, only one out of the four roots yield feasible plots without any singularity and discontinuity. This single root generates 2-D and 4-D profiles that are in good agreement with the ones analytically developed for similar plasma parameters. The two sets of equivalent results along with the various comparable 2-D and 4-D profiles finally describe the SPR behavior of the system.

The SPR dynamics is analyzed with local linear perturbation theory. The corresponding sets of governing equations are given as follows

The continuity equation for electrons (ions)

$$\partial_t n_{e(i)} + \vec{\nabla}.\left(n_{e(i)} \vec{v}_{e(i)}\right) = 0. \tag{1}$$

The total number of particles in a definite volume varies only upon changes in the particle flux across the surfaces enclosing the volume [23].

The momentum equation for electrons (ions) is given by

$$m_{e(i)} n_{e(i)} [\partial_t \vec{v}_{e(i)} + (\vec{v}_{e(i)} \cdot \vec{\nabla}) \vec{v}_{e(i)}] = -\vec{\nabla} P \mp e n_{e(i)} \vec{E} + \eta (\nabla^2 \vec{v}_{e(i)}) + \left(\zeta + \frac{\eta}{3}\right) \vec{\nabla}(\vec{\nabla} \cdot \vec{v}_{e(i)}) + \vec{p}_{ei(ie)}. \quad (2)$$

The net force exerted on the charges is expressed in terms of the individual parametric forces exerted on them, i.e., due to the pressure gradient ($\vec{\nabla}P$), electric field ($\vec{E}$), spatial velocity variation ($\nabla^2 \vec{v}_{e(i)}$), and momentum gain of electronic fluid due to collision with ions ($\vec{p}_{ei(ie)}$), and vice versa [23]. Poisson equation yields the electrostatic potential distribution in terms of the charge density in the generic notations [4] reads as

$$\nabla^2 \phi = 4\pi e (n_e - n_i). \quad (3)$$

Here, $n_{e(i)}, \vec{v}_{e(i)}, m_{e(i)}, P, e, \vec{E}, \zeta, \eta$, and $\vec{p}_{ei(ie)}$ denote the charge number density, velocity, mass of electrons (ions), thermal pressure, electronic charge, local electric field, bulk viscosity, shear viscosity in the medium, and momentum gain due to plasma constituent collisions in the medium, respectively.

### III. STABILITY ANALYSIS

The local stability analysis of the considered sheath-plasma system is performed with the standard technique of linear normal mode analysis in spherically symmetric geometry. Accordingly, the relevant physical variables ($F$) are linearly perturbed ($F_1$) with respect to their corresponding hydrostatic homogenous equilibrium values ($F_o$), presented symbolically [23] as

$$F(r,t) = F_o + F_1(r,t) = F_o + F_{10}\left(\frac{1}{r}\right)\exp(-i(\omega t - kr)), \quad (4)$$
$$F = [n \quad v \quad \phi]^T, \quad (5)$$
$$F_o = [n_o \quad 0 \quad 0]^T, \quad (6)$$
$$F_1 = [n_1 \quad v_1 \quad \phi_1]^T. \quad (7)$$

Inside the IFB the charged particle velocity is usually constant, and the electric field is zero. However, there can be sudden changes due to electron or ion bunching or their transits through the IFB leading to a locally nonzero electric field, especially on the edges. It is noteworthy that, due to the geometrical effects, the spherical IFB demands a $1/r$-factor dependency in the perturbations of the relevant plasma fluid parameters ($v_{e(i)}$ and $n_{e(i)}$), as clearly shown in Eq. (4). The $1/r$-dependency is simply dropped out for any planar geometry and the only form of perturbation is the usual plane-wave form, given as $F_{10} \exp[-i(\omega t - kr)]$ [24,25]. It enables us to transform our physical model from the defined spherical coordination space ($r, t$) to the Fourier wave space ($k, \omega$) with the linear differential operator equivalence relationships, defined as $\partial_t (\equiv -i\omega), \nabla (= \nabla_r \equiv (ik - 1/r)), \nabla^2 (= \nabla_r^2 \equiv 2/r^2 - $ $k^2 - 2ik/r$), and so forth [26]. The terms $k$ and $\omega$ are wavenumber and angular frequency of the perturbation, respectively.

We use the standard linear perturbation formalism, as shown in Eqs. (4)-(7), over Eqs. (1)-(3) for the electron (ion) dynamics. Apart from that, the following substitution for the equilibrium number density, $n_{e(i)o} = n_c \exp(-\gamma r^2) = n_o$ is applied in accordance with experimental data [13]. Here, $n_c$ is the constant electron (ion) number density at the center of the IFB. The density inside an IFB normally follows a Gaussian shape with $\gamma = 1/2\sigma^2$; where, $\sigma$ is the standard deviation [13]. Accordingly, Eq. (1) can be employed to describe the perturbed electronic (ionic) dynamics given with all usual notations [1] as

$$-i\omega n_{e(i)1} + n_{e(i)o}\left(ik + \frac{1}{r}\right)v_{e(i)1} - 2\gamma r n_c \exp(-\gamma r^2) v_{e(i)1} = 0. \tag{8}$$

We simplify Eq. (8) in terms of $n_{e(i)1}$ and $v_{e(i)1}$ across LHS and RHS, as written below

$$n_{e(i)1} = \left(\frac{1}{i\omega}\right)\left(ik + \frac{1}{r} - 2\gamma r\right)n_o v_{e(i)1}. \tag{9}$$

We apply the same formalism in Eq. (2) for the electron (ion) dynamics in the plasma governed by the isothermal classical equation of state. The expressions for the thermal pressure, $P = n_o k_B T_e$, and for the linear momentum gain due to the interparticle (ambipolar) collisional dynamics, $p_{ei(ie)} = [\pi(en_{e(i)o})^2 m_e^{1/2}/(4\pi\epsilon_o)^2(k_B T_e)^{3/2}] v_{e(i)1}$, are adopted from the literature [27]. Further, $p_{ei(ie)}$ is used as $p_o n_o^2 v_{e(i)}$ for the sake of brevity. Accordingly, Eq. (2) with these substitutions reads as

$$-i\omega m_{e(i)} n_{e(i)o} v_{e(i)1} = -\left(ik - \frac{1}{r}\right)k_B T_e n_{e(i)1} \mp \left(ik - \frac{1}{r}\right)en_{e(i)o}\phi_1 - \left(\zeta + \frac{\eta}{3}\right)\left(k^2 + \frac{2}{r^2}\right) - \eta k^2 v_{e(i)1} + p_o n_o^2 v_{e(i)1}. \tag{10}$$

We simplify Eq. (10) in terms of $v_{e(i)1}$, $n_{e(i)1}$, and $\phi_1$ as follows

$$v_{e(i)1} = \frac{k_B T_e\left(ik - \frac{1}{r}\right)n_{e(i)1} \pm en_o\left(ik - \frac{1}{r}\right)\phi_1}{i\omega m_{e(i)} n_o - \left(\left(\zeta + \frac{4\eta}{3}\right)k^2 + \left(\zeta + \frac{\eta}{3}\right)\frac{2}{r^2}\right) + p_o n_o^2}. \tag{11}$$

Using the coefficients of viscosity terms, $\xi = (\zeta + 4\eta/3)$ and $\beta = (\zeta + \eta/3)$ as the effective generalized fluid viscosity and compound fluid viscosity, respectively, the Eq. (11) is modified as

$$v_{e(i)1} = \frac{k_B T_e\left(ik - \frac{1}{r}\right)n_{e(i)1} \pm en_o\left(ik - \frac{1}{r}\right)\phi_1}{i\omega m_{e(i)} n_o - \left(\xi k^2 + \frac{2\beta}{r^2}\right) + p_o n_o^2}. \tag{12}$$

Further, substituting Eq. (12) in Eq. (9) for $v_{e(i)1}$ and after some rearrangement, we have

$$n_{e(i)1} = \frac{\pm en_o^2\left(ik - \frac{1}{r}\right)\left(ik + \frac{1}{r} - 2\gamma r\right)\phi_1}{\left[i\omega\left\{i\omega m_{e(i)} n_o - \left(\xi k^2 + \frac{2\beta}{r^2}\right) + p_o n_o^2\right\} - n_o k_B T_e\left(ik - \frac{1}{r}\right)\left(ik + \frac{1}{r} - 2\gamma r\right)\right]}. \tag{13}$$

We finally apply the same perturbation scheme (Eqs. (4)-(7)) for $\phi_1$ in Eq. (3) with the application of the Laplacian operator in spherical coordinates and arrive at the following equation

$$-k^2 \phi_1 = 4\pi e (n_{e1} - n_{i1}). \tag{14}$$

After substituting the expression for $n_{e1}$ and for $n_{i1}$ from Eq. (13) in Eq. (14), one obtains

$$-k^2 \phi_1 = 4\pi e^2 n_o^2 \left(ik - \frac{1}{r}\right)\left(ik + \frac{1}{r} - 2\gamma r\right)\left[\frac{1}{i\omega\{i\omega m_e n_o - \left(\xi k^2 + \frac{2\beta}{r^2}\right) + p_o n_o^2\} - n_o k_B T_e \left(ik - \frac{1}{r}\right)\left(ik + \frac{1}{r} - 2\gamma r\right)} - \frac{1}{i\omega\{i\omega m_i n_o - \left(\xi k^2 + \frac{2\beta}{r^2}\right) + p_o n_o^2\} - n_o k_B T_e \left(ik - \frac{1}{r}\right)\left(ik + \frac{1}{r} - 2\gamma r\right)}\right] \phi_1. \tag{15}$$

Simplifying Eq. (15) further by arranging the two denominators on LHS, we find the quartic equation given below

$$\left[-\omega^2 m_e n_o - i\omega\left(\xi k^2 + \frac{2\beta}{r^2}\right) + i\omega p_o n_o^2 - n_o k_B T_e \left(ik - \frac{1}{r}\right)\left(ik + \frac{1}{r} - 2\gamma r\right)\right]\left[-\omega^2 m_i n_o - i\omega\left(\xi k^2 + \frac{2\beta}{r^2}\right) + i\omega p_o n_o^2 - n_o k_B T_e \left(ik - \frac{1}{r}\right)\left(ik + \frac{1}{r} - 2\gamma r\right)\right] = 4\pi e^2 \frac{n_o^3}{k^2}\left(ik - \frac{1}{r}\right)\left(ik + \frac{1}{r} - 2\gamma r\right)(m_i - m_e)\omega^2. \tag{16}$$

Since, $m_i \gg m_e$, hence, we further use $m_i \pm m_e \approx m_i$. The modified form of Eq. (16) is

$$\omega^4 m_e m_i n_o^2 + i\omega^3 n_o m_i \left(\xi k^2 + \frac{2\beta}{r^2}\right) - i\omega^3 m_i n_o^3 p_o + \omega^2 m_i n_o^2 k_B T_e \left(ik - \frac{1}{r}\right)\left(ik + \frac{1}{r} - 2\gamma r\right) - \omega^2 \left(\xi k^2 + \frac{2\beta}{r^2}\right)^2 + 2\omega^2 p_o n_o^2 \left(\xi k^2 + \frac{2\beta}{r^2}\right) - \omega^2 p_o^2 n_o^4 + 2i\omega n_o k_B T_e \left(\xi k^2 + \frac{2\beta}{r^2}\right)\left(ik - \frac{1}{r}\right)\left(ik + \frac{1}{r} - 2\gamma r\right)\right) - 2i\omega p_o n_o^3 k_B T_e \left(ik - \frac{1}{r}\right)\left(ik + \frac{1}{r} - 2\gamma r\right) - 4\pi m_i \omega^2 e^2 \frac{n_o^3}{k^2}\left(ik - \frac{1}{r}\right)\left(ik + \frac{1}{r} - 2\gamma r\right) + (n_o k_B T_e)^2 \left(ik - \frac{1}{r}\right)^2 \left(ik + \frac{1}{r} - 2\gamma r\right)^2 = 0. \tag{17}$$

Dividing all the terms with $m_e m_i n_o^2$ (coefficient of $\omega^4$), the Eq. (17) simplifies as

$$\omega^4 + i\omega^3 \left(\frac{1}{m_e n_o}\right)\left(\xi k^2 + \frac{2\beta}{r^2}\right) - i\omega^3 \left(\frac{m_i p_o}{m_e}\right) n_o + \omega^2 \left\{\frac{k_B T_e}{m_e}\left(ik - \frac{1}{r}\right)\left(ik + \frac{1}{r} - 2\gamma r\right) - \left(\frac{1}{m_e m_i n_o^2}\right)\left(\xi k^2 + \frac{2\beta}{r^2}\right)^2 + \left(\frac{2p_o}{m_e m_i}\right)\left(\xi k^2 + \frac{2\beta}{r^2}\right) - \frac{p_o n_o^2}{m_e m_i} - \left(\frac{4\pi e^2}{m_e}\right)\frac{n_o}{k^2}\left(ik - \frac{1}{r}\right)\left(ik + \frac{1}{r} - 2\gamma r\right)\right\} + i\omega \left\{\left(\frac{2k_B T_e}{m_e m_i n_o}\right)\left(\xi k^2 + \frac{2\beta}{r^2}\right)\left(ik - \frac{1}{r}\right)\left(ik + \frac{1}{r} - 2\gamma r\right)\right) - \left(\frac{2p_o n_o k_B T_e}{m_e m_i}\right)\left(ik - \frac{1}{r}\right)\left(ik + \frac{1}{r} - 2\gamma r\right)\right\} + \frac{(k_B T_e)^2}{m_e m_i}\left(ik - \frac{1}{r}\right)^2 \left(ik + \frac{1}{r} - 2\gamma r\right)^2 = 0. \tag{18}$$

To assume the magnitudes of the involved parameters for simplification of Eq. (18), we consider an experimentally reported hydrogen plasma system [5]. This experimental work does not report any excitation of SPR in the studied IFB system. However, this still helps in anticipating the magnitudes of the involved parameters. These typical experimental values help in solving the DR analytically. The reported experimental values of the involved parameters are given as follows [5]

$m_e = 9.1 \times 10^{-31}$ kg, $m_i = 1.67 \times 10^{-27}$ kg, $e = 1.6 \times 10^{-19}$ C, $n_c = 10^{14} - 10^{17}$ m$^{-3}$, $n_o = n_c \exp(-\gamma r^2)$, $k_B T_e = 3.2 \times 10^{-19}$ J, $\xi \approx \beta \approx 10^{-5}$ N s m$^{-2}$, $\gamma = 510$ m$^{-2}$, $p_o = 2.42 \times 10^{-42}$ kg m$^3$/C$^2$s.

These parametric values are specific to only one experimental arrangement. However, from the general solution of the DR, its specific form can be figured out for any other systems as well. Using the following parameters in Eq. (18) and ignoring the second coefficient of $\omega^3$ due to its small magnitude ($\sim 10^{-39}$) in comparison to the first coefficient ($\sim 10^{10}$), we have

$$\omega^4 + \left[1.18i \times 10^{10} \exp(\gamma r^2)\left(k^2 + \frac{2}{r^2}\right)\right]\omega^3 - \left[3 \times 10^{16} \exp(2\gamma r^2)\left(k^2 + \frac{2}{r^2}\right)^2\right]\omega^2 + \left[2.27i \times 10^{18} \exp(\gamma r^2)\left(k^2 + \frac{2}{r^2}\right)\left(ik - \frac{1}{r}\right)\left(ik + \frac{1}{r} - 2\gamma r\right)\right]\omega + \left[1.68 \times 10^{19}\left(ik - \frac{1}{r}\right)^2\left(ik + \frac{1}{r} - 2\gamma r\right)^2\right] = 0. \qquad (19)$$

Clearly, Eq. (19) represents a generalized linear quartic DR in a regular full form with a unique set of multi-parametric dispersion coefficients given respectively as

$A = 1,$ (20)
$B = 1.18i \times 10^{10} \exp(\gamma r^2)\left(k^2 + \frac{2}{r^2}\right),$ (21)
$C = -3 \times 10^{16} \exp(2\gamma r^2)\left(k^2 + \frac{2}{r^2}\right),$ (22)
$D = 2.27i \times 10^{18} \exp(\gamma r^2)\left(k^2 + \frac{2}{r^2}\right)\left(ik - \frac{1}{r}\right)\left(ik + \frac{1}{r} - 2\gamma r\right),$ (23)
$E = 1.68 \times 10^{19}\left(ik - \frac{1}{r}\right)^2\left(ik + \frac{1}{r} - 2\gamma r\right)^2.$ (24)

The SPR frequency is expected to be of the order of the electron plasma frequency ($\sim 10^{11-12}$ Hz) [17,18]. Assuming an $\omega$-value of this magnitude, the first three terms ($\omega^4$, $\omega^3$, $\omega^2$, and their respective coefficients) are found to possess equivalent magnitudes ($\sim 10^{40}$). The approximate magnitudes of the $\omega^1$- and $\omega^0$-terms in Eq. (19) are $10^{28}$ and $10^{19}$, respectively. These are far smaller than in its first three terms. Hence, we ignore these two lower-order terms and move forward with the first three higher-order terms in Eq. (19) justifiably. The reduced form of Eq. (19) is now cast as

$$\omega^4 + \left[1.18i \times 10^{10} \exp(\gamma r^2)\left(k^2 + \frac{2}{r^2}\right)\right]\omega^3 - \left[3 \times 10^{16} \exp(2\gamma r^2)\left(k^2 + \frac{2}{r^2}\right)^2\right]\omega^2 = 0. \qquad (25)$$

It is evident that Eq. (25) is in a simpler form against Eq. (19) due to the following DR roots

$\omega_1 = 0,$ (26)

$\omega_2 = 0,$ (27)

$$\omega^2 + \left[1.18i \times 10^{10} \exp(\gamma r^2)\left(k^2 + \frac{2}{r^2}\right)\right]\omega - \left[3 \times 10^{16} \exp(2\gamma r^2)\left(k^2 + \frac{2}{r^2}\right)^2\right] = 0. \tag{28}$$

Solving Eq. (28) using the quadratic formula

$$\omega = \frac{1}{2A}\left[-B \pm \sqrt{B^2 - 4AC}\right]. \tag{29}$$

Here, $A$, $B$, and $C$ are the coefficients of $\omega^2$, $\omega^1$, and $\omega^0$, respectively in Eq. (28).

Using Eq. (29), the two solutions of Eq. (28) are found to be as follows

$$\omega_3 = 0, \tag{30}$$

$$\omega_4 = -1.18i \times 10^{10} \exp(\gamma r^2)\left(k^2 + \frac{2}{r^2}\right). \tag{31}$$

It may be noticed that the expression in Eq. (31) is the first coefficient of $\omega^3$ in the DR (Eqs. 18-19). The expression is independent of the ion mass, which is a possible variable to different background plasmas. So, Eq. (31) can be used (extrapolated) to find out the general solution for SPR dynamics in any laboratory plasmas. So, the general solution for the SPR in the IFB system is given as

$$\omega_4 = -i\left(\frac{1}{m_e n_c}\right)\left(\xi k^2 + \frac{2\beta}{r^2}\right)\exp(\gamma r^2). \tag{32}$$

The absence of any real component in $\omega = \omega_4$ in Eq. (32) indicates that this SPI branch is not propagatory. The negative imaginary feature of $\omega_4$ points at the stable (with certain conditions) and standing wave-like nature of the SPR. It means that the instability gets reflected from the sheath region. So, the electron plasma frequency in the sheath ($\omega_{spe}$) must be higher than the instability frequency ($\omega_4$), else it will leak through the IFB anode. Consequently, the IFB does not act as a cavity resonator. In addition, the DR (Eq. 18) is also solved numerically using Mathematica with the same set of parametric values. The detailed analytic calculations leading to the derived numerical solutions are skipped for spatiotemporal limitations. Here, the analytical and numerical DR roots are conjointly illustrated for a common multi-parametric comparison graphically. It is observed through comparison that both analytical and numerical figures are in good agreement with each other. This also proves the accuracy of the obtained analytical solution (Eq. 32) obtained with some reasonable approximations.

### IV. RESULTS AND DISCUSSIONS

The dynamics of the excited SPR inside an IFB is studied with a linear perturbation formalism in laboratory spatiotemporal scales in spherical geometry. This linear formalism is applicable under the assumption of small amplitude perturbations [28]. The assumed spherical symmetry makes the polar and azimuthal coordinates within the IFB redundant in the SPR description. This symmetry simplifies the linear analysis by reducing the IFB dynamics into a one-dimensional (radial) problem.

The linear first-order perturbation analysis transforms the IFB model to a generalized linear quartic DR (Eq. 19) with multiparametric coefficients (Eqs. 20-24). To evaluate the contribution of each of the terms along with their coefficients, the values of all the involved constants, such as $e$, $m_{e(i)}$, $p$, $n_o$, and $k_B T_e$ are applied from experimentally reported hydrogen plasma IFB system [5]).

It may be observed that the relevant solution (Eq. 31) of the DR i.e., $-1.18i \times 10^{10}(k^2 + 2/r^2)$ is merely '−1' times the $\omega^3$-coefficient in it (Eqs. 18-19). The terms, such as $n_o$, $k^2$, and $r^2$ do not vary with the background plasma but varies only with the IFB plasma parameters. During the SPR, the two parameters $k$ and $r$ are mathematically linked through the resonance condition $rk = 1/4\pi$. The viscosity coefficients, $\xi$ and $\beta$ are specific to the host plasma used for exciting the SPR. Therefore, it may be inferred from Eq. (31) that the common mode of the SPR in any IFB plasma system is given as $\omega = -i(1/m_e n_c)(\xi k^2 + 2\beta/r^2)\exp(\gamma r^2)$. Applying the experimentally reported values of the various relevant parameters, the SPR mode can be evaluated for that specific IFB plasma system.

The colormap of the SPR dynamics (via Eq. (19)) reveals diverse multi-parametric properties. These analyses comprise of results obtained from the analytical solution (only $\omega_i$) as well as the numerical solution (both $\omega_r$ and $\omega_i$). It is important to add that $\omega_r > 0$ and $\omega_r < 0$ denote the propagating and evanescent nature of the perturbations [29], respectively. Whereas $\omega_i > 0$ and $\omega_i < 0$, on the other hand, denote the growth and decay of the perturbations [29]. In the case of $\omega_r = 0$, the SPR is a pure standing wave, i.e., the nodes and antinodes of the wave formed within the IFB remain spatially invariant over time. For a direct comparison of the results from two different methods (analytical and numerical), the plots are arranged correspondingly. The various 2-D plots (Figs. 2-6, Figs. 10-13) demonstrate the $\omega_{i(r)}$ variation with respect to $r$ or $k$ for different $\xi(\approx \beta)$, $n_c$, and $\gamma$-values. Similarly, the 4-D colormaps show the variation of $\omega_{i(r)}$ with respect to $n_o(r)$, $\chi$ ($= \xi r^2 + 2\beta/r^2$), $r$ or $k$ at the resonance condition $[rk = (1/4\pi)]$.

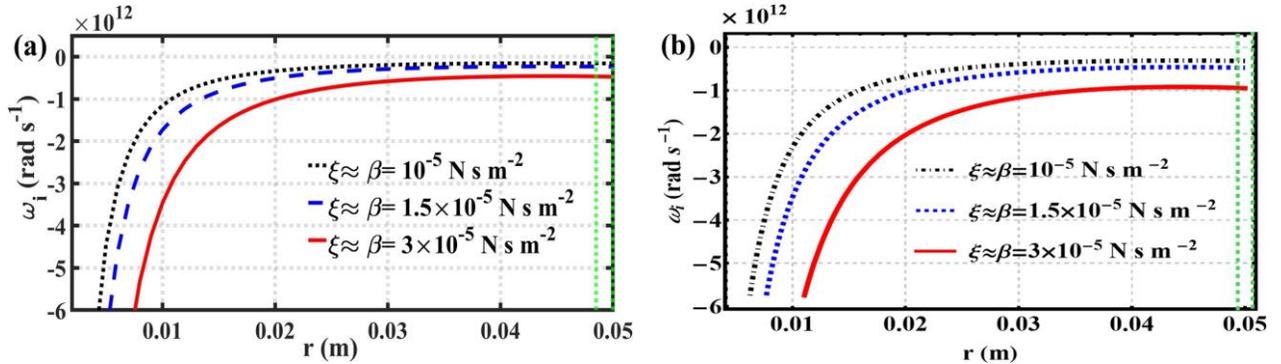

Figure 2: Profile of the spatial variation of the imaginary frequency part ($\omega_i$) of the DR obtained (a) analytically and (b) numerically. The distinct lines link to $\xi(\approx \beta) = 10^{-5}$ N s m$^{-2}$ (black dotted line), $1.5 \times 10^{-5}$ N s m$^{-2}$ (blue dashed line), $3 \times 10^{-5}$ N s m$^{-2}$ (red solid line), respectively. The two externally drawn parallel green dotted lines denote the IFB sheath regions in both the subplots.

The highest SPI amplitude ($\propto \omega_i$) occurs close to the sheath region inside the IFB, where the plasma density and, thus, viscosity ($\chi \propto n_{e(i)}/\nu_{e(i)}$) [30] are at their lowest magnitudes. The $\omega_i \approx 0$ beside the sheath indicates the inability of the IFB system to damp the introduced perturbation in that region. The spatial variation of $\omega_i$ (Figs. 2-3) across the IFB implies a changing response of the IFB plasma towards the SPI. The peripheral IFB sheath region is favorable to the SPI (with higher $\omega_i$-value), whereas the central denser region shows a rapid decrease in the $\omega_i$-value. This also replicates an evanescent wave with amplitude, $F_{10}\exp(\omega_i t)$ gradually decaying towards the center of the IFB. Since the number density $[n_o = n_c \exp(-\gamma r^2)]$ has its maximum ($n_c$) at the IFB center ($r = 0$), it clearly proves that increasing $n_o$ damps the SPI amplitude.

The apparent difference in the radial variation of the $\omega_i$-profiles relative to the IFB center among the analytical (Fig. 2(a)) and numerical (Fig. 2(b)) subplots may be explained by a minor influence from some of the omitted terms while solving the DR (Eq. (19)) analytically. Consequently, the curves in analytical plots start growing a bit closer to the IFB center because the electron density increases there, as discussed below. Since all the possible terms are considered in the numerical solution, they make some small contributions to the effective plasma density. Thus, the instability grows a bit further away from the center.

The physics behind the SPR excitation may be explained with the inductor (L) and capacitor (C)-like behavior of the field free IFB plasma and the sheath, respectively. As the interior of the IFB is field free and consists mainly of cold electrons, it behaves inductively and can become resonant to its surrounding sheath, which essentially acts as a capacitor [17,18]. As discussed in [17], this LC resonant circuit exhibits an unstable antiresonance, that depends only weakly on the IFB electrode bias. The instability occurs due to negative RF sheath resistance during the SPR, when the electrons also start to behave inertially along with the ions. The SPR occurs most efficiently in the ideal case, when $\omega^2/\omega_p^2 \approx 1$. Furthermore, the instability is strongly damped at the plasma potential.

The electron density increases towards the center of the IFB and so does the plasma frequency. Hence, it approaches the limit $\omega/\omega_p \to 0$. In this extreme case the resonant part of the wave is extinguished while the antiresonance that drives the SPI is strongly damped, a result that also corroborates the experimental findings [17]. The results of our model, which indicate this behavior, are also depicted in the figures below.

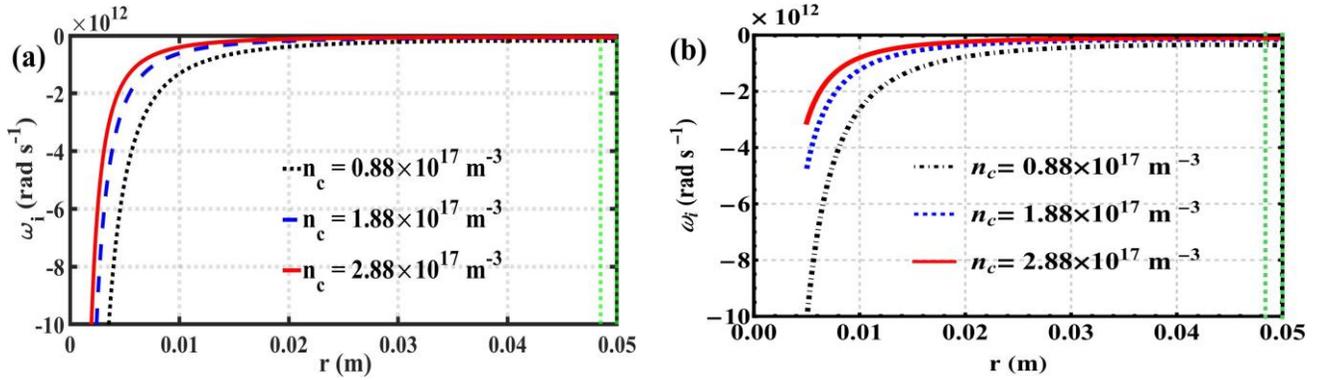

Figure 3: Same as Fig. 2, but for various $n_c$-values. The distinct lines correspond to $n_c = 0.88 \times 10^{17}$ m$^{-3}$ (black dotted line), $1.88 \times 10^{17}$ m$^{-3}$ (blue dashed line), $2.88 \times 10^{17}$ m$^{-3}$ (red solid line), respectively.

It may be concluded from Fig. 3 that the IFB center is most stable against any introduced SPR perturbation. The SPR, which originates in the sheath region is also strongest there and dies out while approaching the center. The wavelength of the acoustic wave produced due to SPR is equivalent to the diameter of the IFB anode. This indicates that the IFB can act as a cavity resonator holding an SPR induced standing wave. It may also be pointed that there is no qualitative variation in $\omega_i$ noticed against the central density ($n_c$) apart from the meagre change in $\omega_i$-magnitude.

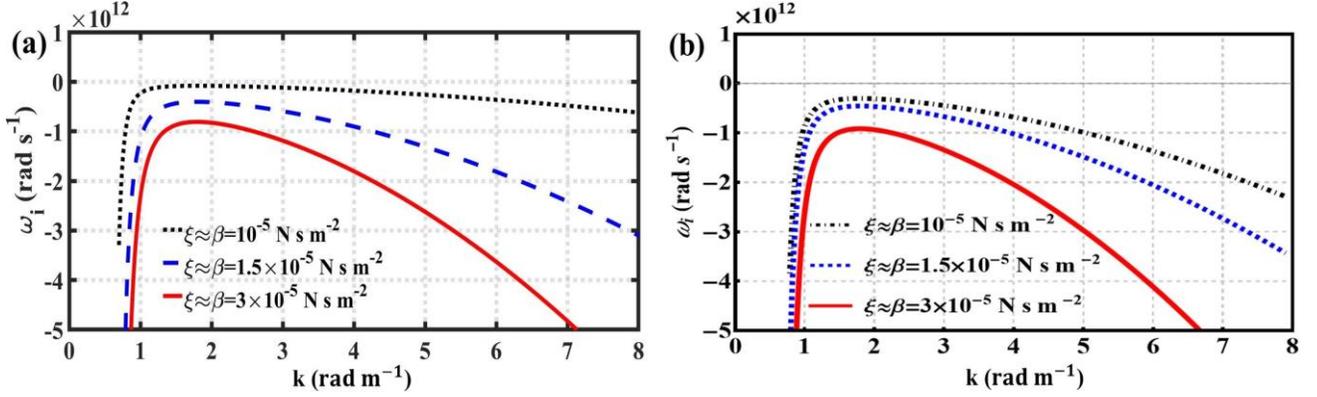

Figure 4: Same as Fig. 2, but in $k$-space.

Applying the IFB resonance condition of $r = 1/4\pi k$, the final solution of the DR (Eq. (31)) is expressed solely in terms of wave number $k$ through both the analytical (Fig. 4(a)) and numerical (Fig. 4(b)) profiles. The modified form which reads $\omega_i = -316.51 k^2 (1/m_e n_c) \exp(\gamma/16\pi^2 k^2)$, is plotted for various values of $\xi (\approx \beta) = 10^{-5}$ N s m$^{-2}$ (black dotted line), $1.5 \times 10^{-5}$ N s m$^{-2}$ (blue dashed line), and $3 \times 10^{-5}$ N s m$^{-2}$ (red solid line)). The SPI shows a sudden rise with respect to $k$ with subsequent fall at higher $k$-values. The $\omega_i$-variation in the $k$-space corresponds to Fig. 2 for $r$.

The $\omega_i$-variation in the $k$-space represents a growth (or decay) rate of the SPI for different wavelengths or frequencies of the applied perturbation. It also means that the effective plasma behavior differs for different wavelengths of perturbations. It is seen quite often that the growth (or decay) rate of the instability does not have a linear correlation with $k$. The $k$-value corresponding to the maxima of $\omega_i$-variation may be depicted as the resonance point where the perturbation grows most rapidly.

Moreover, when looking at the instability frequency in the $k$-space (Fig. 4), one sees a distinctive maximum in $\omega_i$ in this domain. The reason for this becomes evident from a closer inspection of the DR solution, Eq. (32). Keeping in mind that the wavenumber $k$ is proportional to $1/r$, one can see that in the case of small $k$ (or large $r$) the exponential term in Eq. (32) becomes dominant while the prefactor vanishes. However, the exponential term goes to 1 for $r \to 1$, which leads to a peak in the $\omega_i(k)$-function close to the IFB center.

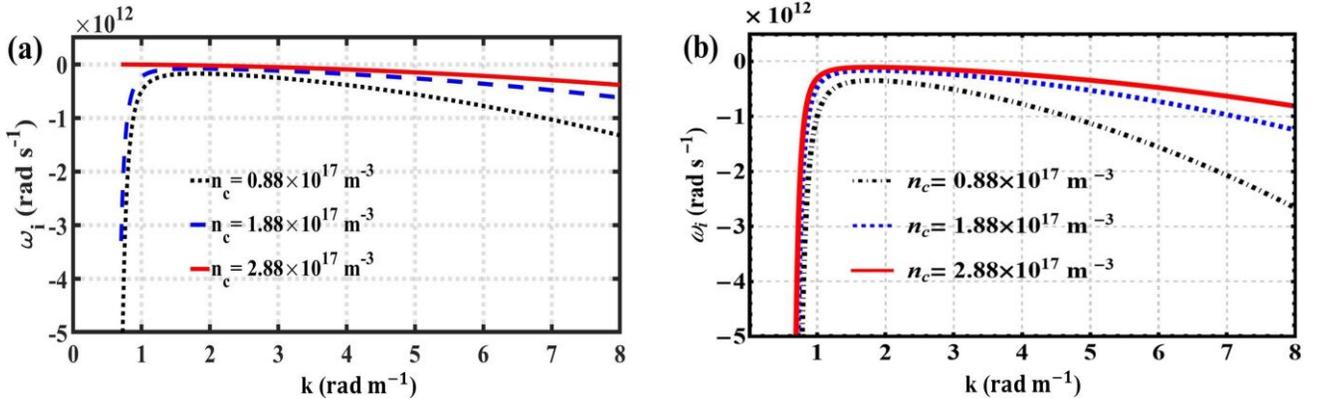

Figure 5: Same as Fig. 3, but in the $k$-space.

It is noteworthy that a good agreement between the analytical and numerical results proves the semi-analytic accuracy and reliability of the proposed SPR calculation scheme. It is quite apparent that $n_c$ does not have any qualitative influence on $\omega_i$, as also observed earlier. However, the exponential drop in the plasma density towards the edge of the IFB influences the oscillation considerably. The corresponding $\omega_i$ curves for three different $n_c$-magnitudes at smaller $k$-values are indistinctly merged. The basic physics behind such $\omega_i$-variations remains the same as already explained before in Fig. 4. The observed $\omega_i$-maxima occurs at $k \approx 1.4$ rad m$^{-1}$ to 1.6 rad m$^{-1}$, which corresponds to, from $r = 5.6 \times 10^{-2}$ m to $4.9 \times 10^{-2}$ m, perfectly covering the sheath region ($r \approx 5 \times 10^{-2}$ m or 5 cm) with the resonance condition ($rk = 1/4\pi$). The oscillation amplitude is maximum beside the sheath region due to the SPR, like an antinode in a standing wave. Therefore, the specific $r$ corresponding to the $k$ where the $\omega_i$-maxima occurs, holds the antinode forming due to the SPR excited inside the IFB.

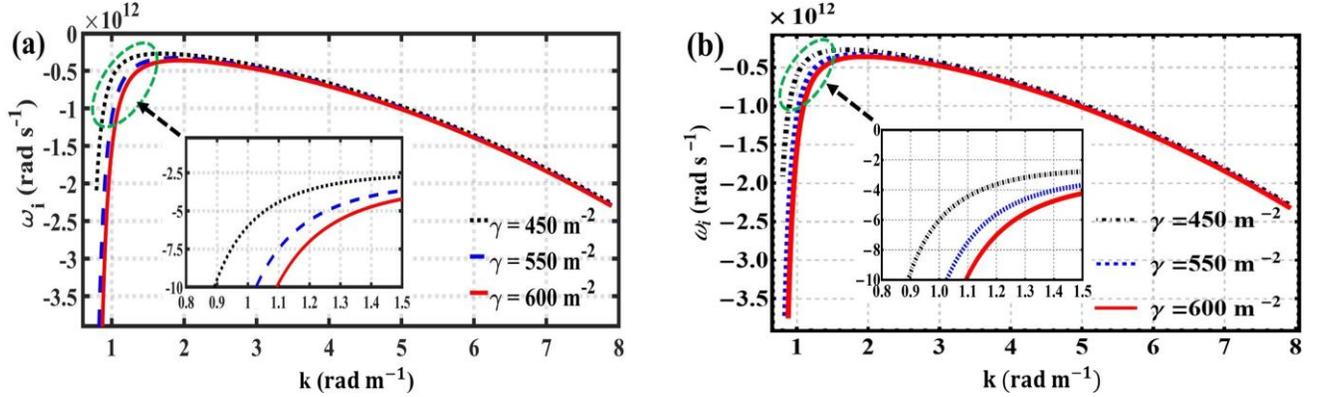

Figure 6: Same as Fig. 4, but for various $\gamma$-values. The different lines link to $\gamma = 450$ m$^{-2}$ (black dotted line), $\gamma = 550$ m$^{-2}$ (blue dashed line), $\gamma = 600$ m$^{-2}$ (red solid line), respectively.

Fig. 6 shows the $\omega_i$-variation against $k$ for different values of $\gamma$. The analytical (Fig. 6(a)) and numerical (Fig. 6(b)) plots are in good agreement for all values of $k$. Whereas, for larger values of $k$ the plots are congruent with slight variation in magnitudes. The experimentally reported value of $\gamma$ for hydrogen plasma is nearly $510$ m$^{-2}$ [13]. The $\omega_i$-magnitude, while lower than $\omega_{spe}$ for $\gamma \approx 510$ m$^{-2}$, shows that the SPR remains trapped inside the IFB. This occurs due to its internal reflection from the inside of the sheath forming a standing acoustic wave like structure.

It may be highlighted that greater $\gamma$-value makes the density variation [$n_o \propto n_c \exp(\gamma r^2)$] more pronounced within the IFB. This means that higher $\gamma$-values indicate a higher accumulation of charges in the IFB center, whereas the outer peripheries contain less charges. As a result, the SPI is more attenuated around the IFB center than anywhere else. Therefore, it may be inferred from the plots (Fig. 6) that $\gamma$ is a stabilizing parameter against the SPI.

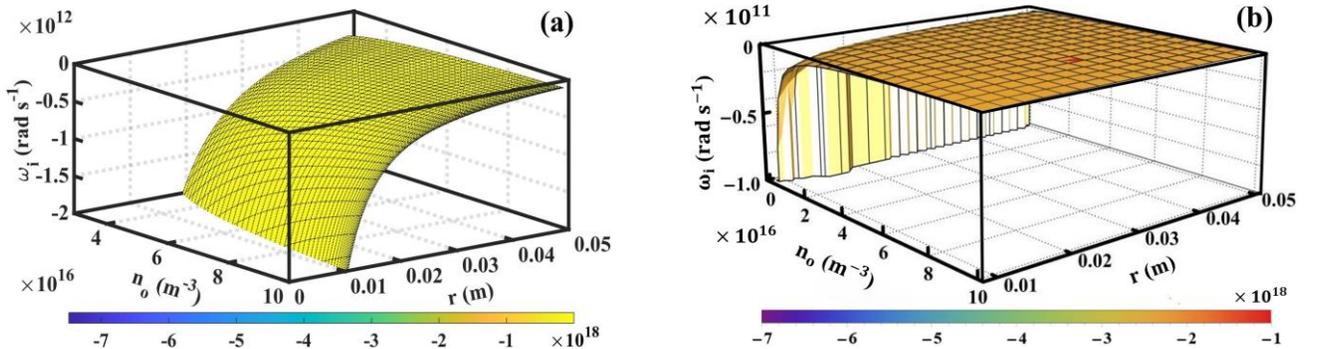

Figure 7: Colormap profile depicting the (a) analytical and (b) numerical $\omega_i$-variation with $r$ and $n_o$ [$= n_c \exp(-\gamma r^2)$].

Fig. 7 shows the collective variation of $\omega_i$ with respect to $r$ and $n_o$. There may be minor discrepancies in the ordinate scaling ($10^{12}$ in Fig. 7(a) (analytical) and $10^{11}$ in Fig. 7(b) (numerical)). It results in axis-parametric deviations originating from the two lowest-order terms expediently ignored in simplifying Eq. (19). It is noticed that $\omega_i$ increases with the increase in $n_o$, but the negative growth indicates the damping of the SPR. The large negative $\omega_i$-magnitude at largest $r$ and smallest $n_o$-values replicates that of 2-D profiles (Fig. 3) and proves that the SPR oscillation is internally reflected by the IFB sheath boundary. The larger $\omega_{spe}$-value does not allow the instability to leak through the IFB boundary. The physics for the colormap (Fig. 7) is same as that explained in Fig. 3 which deals with both the density and radial distance variations.

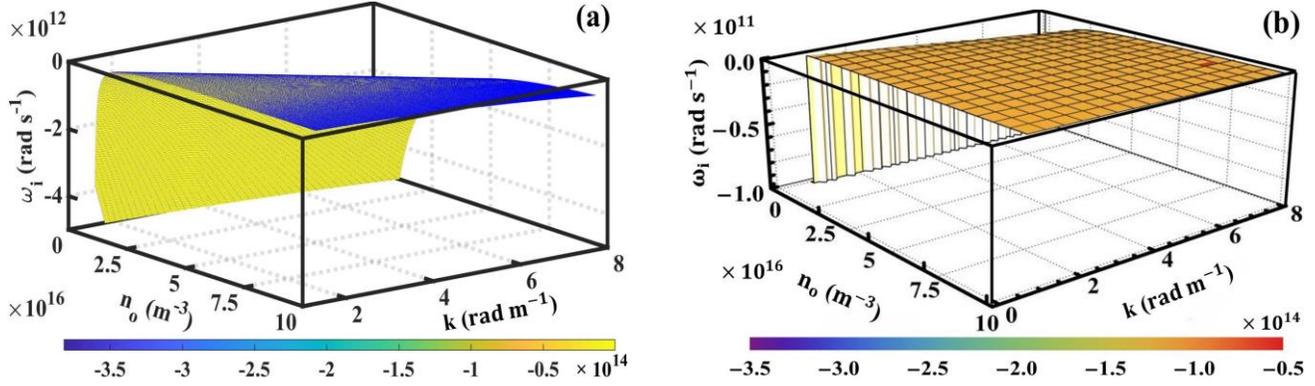

Figure 8: Colormap depicting the (a) analytical and (b) numerical $\omega_i$-variation conjointly with $k$ and $n_o$ [$= n_c \exp(-\gamma r^2)$].

For the SPR, both $r$ and $k$ are related by the expression $r = 1/4\pi k$. Hence, the modified SPR mode can be expressed in terms of $k$ as $\omega_i = -316.51 k^2 i (1/m_e n_c) \exp(\gamma/16\pi^2 k^2)$. When $k$ increases, the $n_o$ also increases, and the $\omega_i$-magnitude reduces. This behavior is opposite compared to the variation in $\omega_i$ with respect to $r$ as shown in Fig. 7, since $r$ and $k$ are inversely proportional to each other. Concisely, $\omega_i$ and $k$ are directly proportional to each other up to certain extent, beyond which they become inversely proportional. The same behavior is also noticed for $\omega_i$ varying with respect to $n_o$. The discrepancies in the graphical ordinate-scaling in Fig. 8(a) as $10^{12}$ (analytical) and in Fig. 8(b) as $10^{11}$ (numerical), their quantitative graphical deviations, are the same as discussed before for Figs. 7(a-b). No such discrepancies are observed in the 2-D plots due to only one independent variable against two variables in the 4-D colormaps.

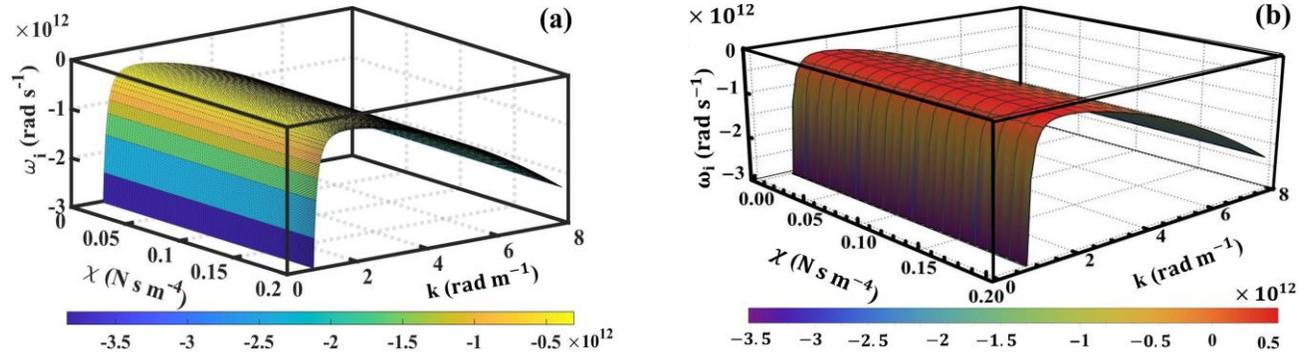

Figure 9: Colormap profile depicting the (a) analytical and (b) numerical resonance $\omega_i$-variation conjointly with $k = 1/4\pi r$ and $\chi\ (= \xi k^2 + 2\beta/r^2)$.

As in Fig. 5, an SPR damping is noticed with the increase in $k$. It should be recalled that there is a variation of fluid viscosity with $k$. Therefore, any coordinate on the colormap is a triangulated outcome of all the three parameters, $\omega_i$, $(\xi k^2 + 2\beta/r^2)$, and $k$. Here, $k$ is the only independent parameter and the rest are dependent singly on $k$. The SPI mode remains trapped inside the IFB and forms a standing wave. It resembles the previously reported experimental data [17,18], thereby corroborating the proposed bifluidic plasma fireball model approach.

After a series of comparisons between the analytical (using MATLAB) and numerical (using Mathematica) graphical results for $\omega_i$-parameter, the numerically obtained nonzero $\omega_r$-parameter is discussed hereafter. It may be reminded that the $\omega_r$ is found to be zero in magnitude in the analytical solution due to its small order of magnitude. Hence, the discussed $\omega_r$-parameter afterwards is purely a yield of the numerical solution of the DR (Eq. (19)).

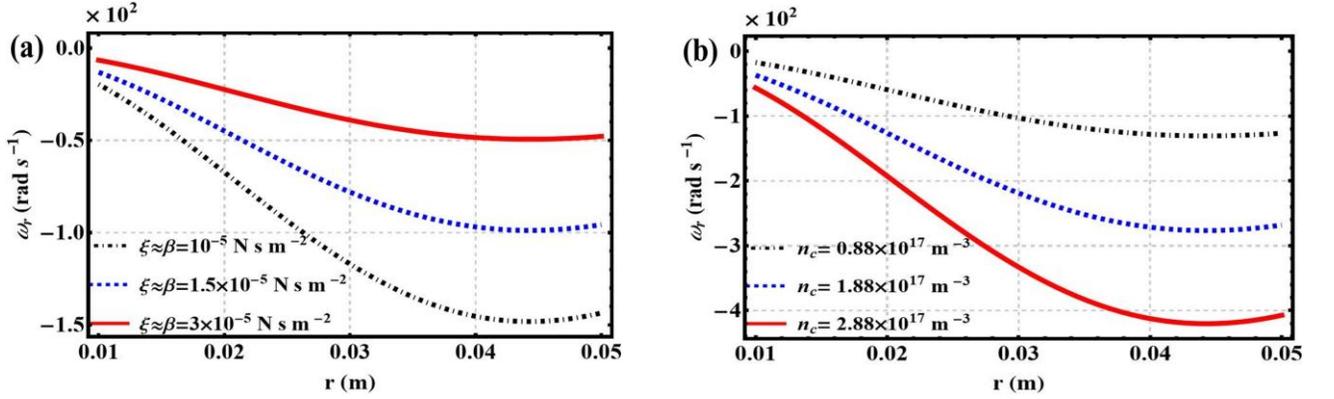

Fig. 10: Variation of $\omega_r$ for various (a) viscosity ($\chi \propto \xi, \beta$) and (b) central density ($n_c$) values with $r$ at the resonance condition ($k = 1/4\pi r$). The distinct lines link to different $\xi\ (\approx \beta)$ and $n_c$ values indicated as (a) $\xi\ (\approx \beta) = 10^{-5}$ N s m$^{-2}$ (black dotted line), $1.5 \times 10^{-5}$ N s m$^{-2}$ (blue dashed line), $3 \times 10^{-5}$ N s m$^{-2}$ (red solid line), and (b) $n_c = 0.88 \times 10^{17}$ m$^{-3}$ (black dotted line), $1.88 \times 10^{17}$ m$^{-3}$ (blue dashed line), $2.88 \times 10^{17}$ m$^{-3}$ (red solid line). The rest are described in the text.

The observed characteristics of subplots in Fig. 10 are exactly opposite to those observed in Figs. 2-3 for the $\omega_i$-parameter. In contrast to Figs. 2-3, larger $n_c$- and smaller $\xi(\approx \beta)$-values are found to yield higher values for $\omega_r$ (Fig. 10). The physics for this variation may again be attributed to the rising number density, thus plasma frequency towards the center, which strongly damps the instability.

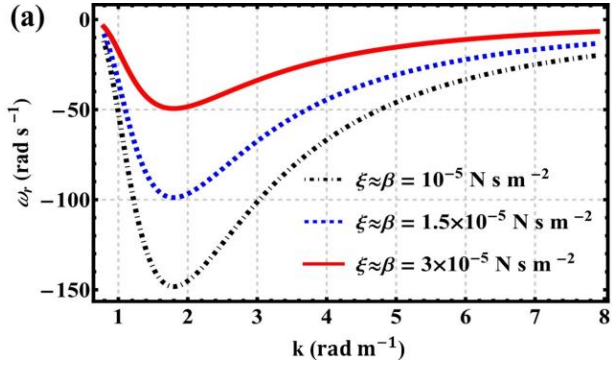
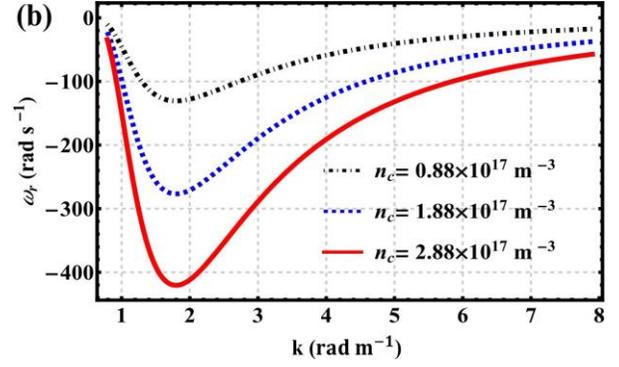

Fig. 11: Same as Fig. 10, but in the $k$-space.

It is seen that Fig. 11 is comparable with Figs. 4-5. The corresponding maxima and minima of $\omega_i$ and $\omega_r$ for two different parameters [$\xi(\approx \beta)$, and $n_c$] form at common $k$-values.

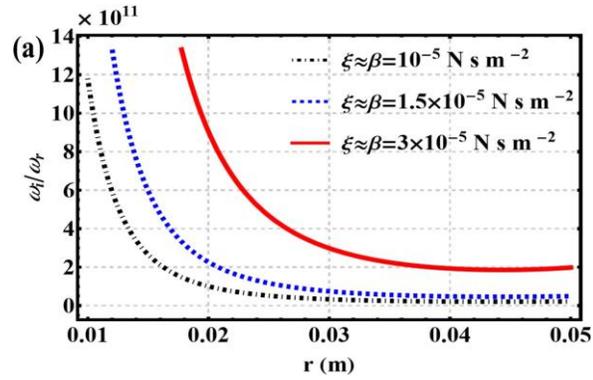
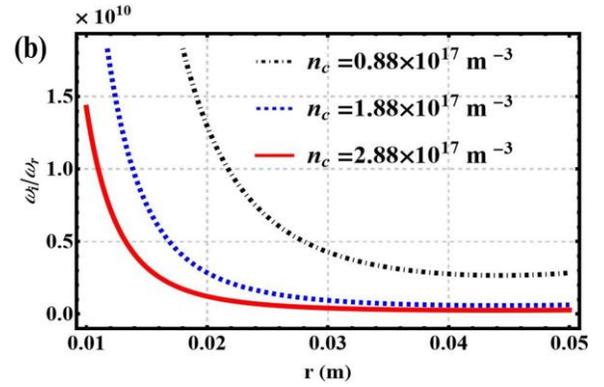

Fig. 12: Comparative profiles of $\omega_i$ and $\omega_r$ with respect to $r$ for different (a) $\xi(\approx \beta)$ and (b) $n_c$-values.

The spatial correlation of $\omega_i$ and $\omega_r$ for different parameters [$\xi(=\beta)$, and $n_c$] proves that the SPI ($\omega_i > 0$) and the evanescent wave formation ($\omega_r < 0$) occur simultaneously, and the latter is ubiquitous in the SPR.

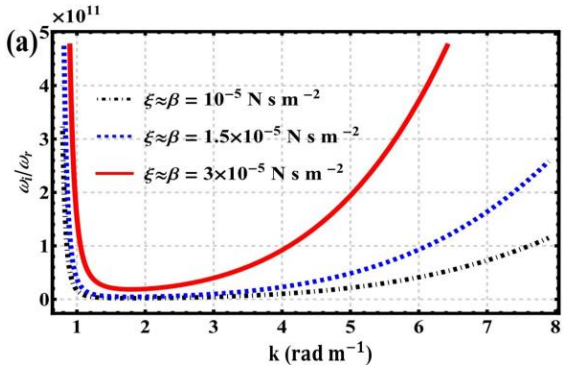
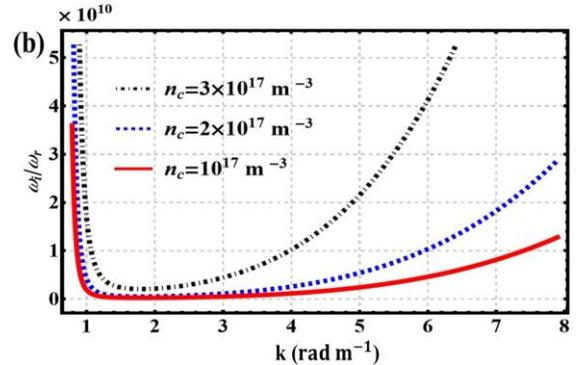

Fig. 13: Same as Fig. 12, but in the $k$-space.

Figure 13 compares the magnitudes of $\omega_i$ (instability or stability) and $\omega_r$ (propagation or evanescence) with respect to $k$ for viscosity and density variations. The smaller $\omega_r$-magnitude yields a larger $\omega_i/\omega_r$-ratio of order~$10^{10}$-$10^{11}$. Larger $\omega_i$ and smaller $\omega_r$ prove the standing wave-like and reflecting SPI-nature at the inner edge of an IFB.

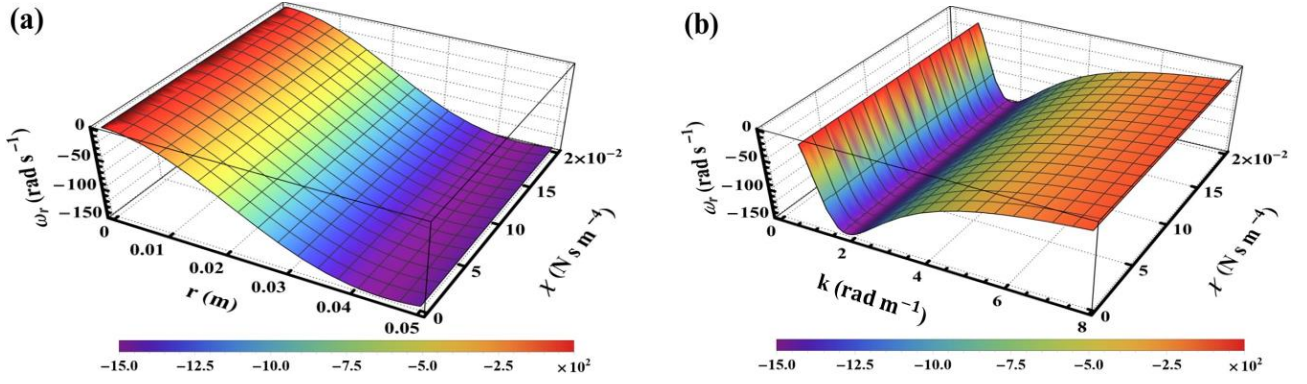

Fig. 14: Colormap showing the $\omega_r$-profile with variation of $\chi$ evolving in (a) $r$ and (b) $k$.

The respective rise and fall in the perturbation expressed through $\omega_i$-variations are due to an increased plasma density close to the IFB center. The lower $\omega_r$-value towards the IFB boundary (sheath region) indicates an evanescent SPI. The $\omega_r$-parameter gradually approaching zero indicates a purely standing wave SPI at the IFB center.

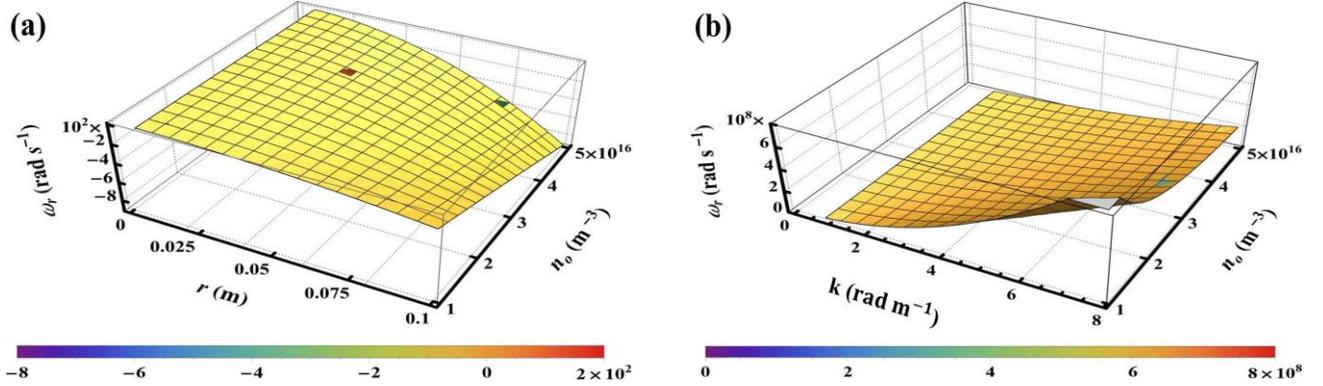

Fig. 15: 4-D Colormap showing the collective variation of $\omega_r$ with respect to $n_o$, (a) $r$, and (b) $k$.

It is noticeable that, Fig. 7 and Fig. 15 illustratively show a similar variational pattern in $\omega_i$ and in $\omega_r$, respectively. It is further seen that all the 2-D line profiles (Figs. 2-6, Figs. 10-13) and 4-D color profiles (Figs. 7-9, Figs. 14-15) from both the analytical and numerical outcomes corroborate each other. It justifies the reliability of our semi-analytical calculation scheme.

The multi-parametric influences on the SPR stability features (Figs. 1-15) are now highlighted in Table 1 for the sake of instant reference of its readers. Besides, a tabular comparison of the analytical and numerical results is also presented in Table 2. It draws a relative contrast between these two sets of results by using two different software tools for graphical analyses. It uses MATLAB for analytical results and Mathematica for numerical ones. The similarities and the dissimilarities between the two sets of graphical results are highlighted in Table 2. Their fine matching expressed through their corresponding plots emboldens the accuracy of the SPR analysis. The difference observed in the plots in terms of parametric spatial variations is due to a few lower magnitude terms ignored while solving the DR (Eq. 19) analytically as mentioned before.

## TABLE I. MULTI-PARAMETRIC SPR STABILITY

| S. No. | Physical parameter | Influence | Figure | Physical remark |
|---|---|---|---|---|
| 1 | Central density ($n_c$) | Weakly damping | Figs. 3, 5, 7, 8 | Density fields act against the acoustic oscillations |
| 2 | Density steepness parameter ($\gamma$) | Damping | Fig. 6 | Higher steepness increases the rate of damping at the IFB center and vice-versa |
| 3 | Effective viscosity ($\xi r^2 + 2\beta/r^2$) | Damping | Figs. 9, 14 | High-viscosity plasmic shells prevent acoustic oscillations |

## TABLE II. ANALYTICAL VS. NUMERICAL RESULTS

| S. No. | Item | Contrast | Figure | Remarks |
|---|---|---|---|---|
| 1 | $\omega_i = \omega_i(r)$ for different $\xi$-values | Similar | Fig. 2 | Two results corroborate |
| 2 | $\omega_i = \omega_i(r)$ for different $n_c$ values | Similar | Fig. 3 | Two results corroborate |
| 3 | $\omega_i = \omega_i(k)$ for different $\xi$-values | Similar | Fig. 4 | Two results corroborate |
| 4 | $\omega_i = \omega_i(k)$ for different $n_c$ values | Similar | Fig. 5 | Two results corroborate |
| 5 | $\omega_i = \omega_i(k)$ for different $\gamma$ values | Similar | Fig. 6 | Two results corroborate |
| 6 | $\omega_r$ vs any other parameter | NA | Figs. 10-15 | $\omega_r \approx 0$ for analytical solution and $\omega_r < 0$ for numerical solution |

## V. CONCLUSIONS

The SPR excitation physics in an IFB system is semi-analytically studied in the viewpoint of linear local perturbative treatment in a spherically symmetric geometry. The polar and azimuthal parametric contributions are justifiably ignored for the sake of mathematical simplicity in the adopted calculation scheme. The basic structuring equations for portraying the electron and ion dynamics are linearized due to the assumed small-scale multi-parametric perturbations. This linearization enables us to ignore the nonlinear (higher-order) terms (multi-parametric cross-coupling effects) in the formalism of the plasma system. The electrostatic Poisson equation for the potential distribution yields the multi-parametric model closure as an intrinsic local model coupling property originating from the constitutive local charge density fields. The perturbed model system gets methodically decoupled into a quartic dispersion relation (Eq. (19)). The analysis theoretically demonstrates the SPR excitation in a reticular IFB in spherical geometry, unlike those found in the experimental arrangements, set up with central solid anodes. Such an SPR instability, studied here semi-analytically, has practically been observed and reported in experiments with a solid anode with typical plasma parameters: $n_e < 5 \times 10^{17}$ m$^{-3}$, $k_B T_e \leq 0.5$ eV, and $P < 3 \times 10^{-4}$ Torr [18]. To the best of our knowledge, no SPR excitation has hitherto been reported experimentally in the literature for an IFB system developed with such hollow and reticular anodes as described above.

The obtained DR (Eq. (19)) is simplified by comparing the magnitudes of the multi-parametric coefficients through parametric substitution from an exemplary hydrogen plasma system [5]. The quartic equation is hence analytically reduced to a quadratic equation. It is noteworthy that among the four roots obtained, only one considerably strong SPR mode is found to survive, with no propagation component ($\omega_r = 0$). It indicates that the SPR mode behaves as standing wave-like patterns in nature.

The quartic DR derived and solved analytically above, is now solved numerically using Mathematica, for analyzing the exact DR roots. This technique yields four different roots with fully distinct characteristics. However, only one out of the four roots (the third root) produce feasible results, which are plotted in this work with respect to $r$ and $k$. The lengthy numerical solutions are not presented here in detail. But the obtained plots from the solutions are used in the SPR analysis. A fair matching of the analytical and numerical solution patterns (for $\omega_i$) proves the reliability of our calculation scheme. The numerically evaluated evanescent ($\omega_r < 0$, spatially constrained at the source) nature of the instability within the IFB sheath is also further validated with the help of obtained colourmaps (Fig. 15). Such evanescent modal features within the IFB originate due to the reflecting behavior of the sheath for the SPR frequencies, subcritical against the sheath plasma frequency, $\omega_{spe}$.

The SPR develops in the sheath region and shows spatial variations of $\omega_i$ across the IFB. The externally biased IFB anode serves as the energy source for the SPI. As the density ($n_o$) and viscosity ($\chi$) increase towards the IFB center from the sheath, the SPR disturbance starts damping. This quasi-stable SPR disturbance here shows a steep damping towards the IFB center after a finite traversal. Moreover, the SPR disturbance cannot leak through the sheath as the plasma frequency in the sheath region ($\omega_{spe}$) is higher than $\omega_i$-magnitude. So, it may be concluded that the IFB acts as a cavity resonator during the excitation of the SPR within it. It is because the IFB functions both as a cavity (shield) for the SPR excitations and as its sheath resonates with the internally trapped IFB plasma.

Along with the SPR research, the fireball model is noticed to have many other applications in both pure and applied fields of active research and development. A few of these applications have been explored for decades, while the rest have been studied quite recently. Some of the focal investigative areas prevailing so far in the IFB context are briefly highlighted as follows:

(i) Antenna signals in spacecrafts: The antennas of the spacecrafts get charged in the astroplasmic environments behaving further as electrodes with floating potential. The possible excitations of the SPR in the antenna replicates with that in laboratory plasmas. Therefore, the laboratory SPR studies could help in antenna signal processing and subsequent applied analysis [17].

(ii) Coronal mass ejection: The Sun could usually be modelled as a huge plasma fireball in the thermonuclear fusion perspective. The magnetic turbulence associated with the solar sheath and subsequent coronal mass ejections (CMEs) could also be therewith studied. It is possible via fireball model formalisms in a scale invariant pathway as herein. These CMEs are interestingly active agents to yield major geomagnetic storms (as observed on the 20[th] of February 2023). It has been reported widely via the Wind/WAVES instrumentation and detection technology [31].

(iii) Junction diode analysis: The depletion region in a junction diode behaves as a plasma sheath. This region functions as a barrier in between the n- and p-type charges against diffusion, and the sheath functions as a similar potential barrier in between the electrode and ambient plasmas [32]. So, the proposed SPR analyses could be broadly useful for similar high-frequency diode-biasing experiments, observational analyses, and realistic applications extensively.

(iv) Sheath diagnosis: The SPR analysis reveals multiple plasma properties of great technological importance. It includes mainly the electron (ion) densities, electrostatic sheath potential, transient (displacement) currents, current-voltage characteristics, sheath electric field penetration and its shielding, ion ejection mechanism, potential relaxation oscillations, and so forth [19]. Such outcomes derived during an SPR study could be broadly useful to diagnose and characterize the involved plasma sheath structures in the said configurations comprehensively.

Apart from the above, the proposed model analyses could bring forward various future possible applicability and scope in more practical plasma environments. Such situations may involve highly nonlinear perturbations, presence of inhomogeneous magnetic fields, non-spherical geometries, nonlocal perturbations, and so forth. The study of SPRs in IFBs yields more physical insights into diverse plasma phenomena. It is shown here how the SPR behaves inside an IFB illustratively. Besides, numerical and analytical outcomes are compared elaborately resulting in reasonable corroboration. It is specifically shown herein that the SPR exhibits the collective excitation of evanescent standing wave patterns inside IFBs, subject to specific multi-parametric ranges. These waves fulfill the SPR condition. They are fully reflected by the plasma sheath, surrounding the IFB, thereby rendering it purely an internal IFB instability mode sourced in the free energy associated with the biased electrode (anode).

## ACKNOWLEDGEMENTS
The active support from colleagues from the Astrophysical Plasma and Nonlinear Dynamics Research Laboratory (APNDRL), Tezpur University, in this manuscript preparation, is thankfully acknowledged. This work is dedicated to late Prof. R. L. Stenzel, University of California, USA, the great plasma physicist and experimentalist, who passed away on the 9$^{th}$ of December 2023. One of the coauthors, J.G., feels honored for being guided and mentored by late Prof. Stenzel.

## REFERENCES
[1] S. Dutta and P. K. Karmakar, *Fireball Sheath Instability*, J. Astrophys. Astr. **43**, 64 (2022).
[2] S. Dutta and P. K. Karmakar, *A Multi-Order Nonlinear Meta-Analysis of Bifluidic Fireball Sheath Fluctuations*, Waves Random and Complex Media, DOI: 10.1080/17455030.2023.2178826 (2023).
[3] G. W. Wetherill and M. O'D C. Alexander, *Meteor and Meteoroid*. Encyclopedia Britannica. https://www.britannica.com/science/meteor
[4] R. L. Stenzel, C. Ionita, and R. Schrittwieser, *Dynamics of Fireballs*, Plasma Sources. Sci. Technol. **17**, 035006 (2008).
[5] J. Gruenwald, J. Reynvaan, and P. Knoll, *Creation and Characterization of Inverted Fireballs in $H_2$ Plasma*, Phys Scr. **2014 T161**, 014006 (2014).
[6] J. Gruenwald, et al. *Further Experiments on Inverted Fireballs*, Proc. 38th EPS Conf. Plasma Phys. (2011).
[7] R. L. Stenzel, et al. *Transit Time Instabilities in an Inverted Fireball. I. Basic Properties*, Physics of Plasmas **18** (1), 012104 (2011).
[8] R. L. Stenzel, J. Gruenwald, B. Fonda, C. Ionita, and R. Schrittwieser, *Transit Time Instabilities in an Inverted Fireball. II. Mode Jumping and Nonlinearities*, Phys. Plasmas **18**, 012105 (2011).
[9] R. L. Stenzel, J. Gruenwald, C. Ionita, and R. Schrittwieser, *Pulsating Fireballs with High-Frequency Sheath-Plasma Instabilities*, Plasma Sources. Sci. Technol. **20**, 045017 (2011).
[10] O. Lehmann, *Gasentladungen in Weiten Gefässen*, Annalen der Physik **312**, 1 pp. 1-28 (1902).

[11] I. Langmuir, *Positive Ion Currents from the Positive Column of Mercury Arcs*. Science **58**, 1502 (1923).

[12] R. L. Stenzel, C. Ionita, and R. Schrittwieser, *Plasma Fireballs*, IEEE Transections on Plasma Science **36**, 4 (2008).

[13] J. Gruenwald, *On the Dispersion Relation of the Transit Time Instability in Inverted Fireballs*, Physics of Plasmas **21**, **0**82109 (2014).

[14] J. Gruenwald and M. Fröhlich, *Coupling of Transit Time Instabilities in Electrostatic Confinement Fusion Devices*. Physics of Plasmas **22**, 070701 (2015).

[15] R. L. Stenzel, et al. *Sheaths and Double Layers with Instabilities*. J. Technol. Space Plasmas **2**, 1 pp. 70-921 (2021).

[16] R. L. Stenzel, J. Gruenwald, C. Ionita, et al., *Electron rich sheath dynamics. I. Transient currents and sheath-plasma instabilities*, Phys. Plasmas **18**, 062112 (2011).

[17] R. L. Stenzel, *Instability of Sheath-Plasma Resonance*, Phys. Rev. Lett. **60**, 8 pp. 704-707 (1988).

[18] R. L. Stenzel, *High-Frequency Instability of the Sheath-plasma Resonance*, Physics of Fluids B: Plasma Physics **1**, 2273 (1989).

[19] R. L. Stenzel, J. Gruenwald, C. Ionita, and R. Schrittwieser, *Electron rich sheath dynamics. II. Sheath ionization and relaxation instabilities*, Phys. Plasmas **18**, 062113 (2011).

[20] J. Gruenwald, J. Reynvaan, and P. Geistlinger, *Letter: Influence of Inhomogeneous Electrode Biasing on the Plasma Parameters of Inverted $H_2$ Fireballs*, Journal of Technological and Space Plasmas, **1**, 1 (2024).

[21] R. L. Stenzel, J. Gruenwald, C. Ionita, and R. Schrittwieser, *Pulsed, Unstable and Magnetized Fireballs*, Plasma Sources. Sci. Technol. **21**, 015012 (2012).

[22] V. Mitra, N. H. Prakash, and I. Solomon, et al., *Mixed Mode Oscillations in Presence of Inverted Fireball in an Excitable DC Glow Discharge Magnetized Plasma*, Phys. Plasmas **24**, 022307 (2017).

[23] F. F. Chen, *Introduction to Plasma Physics and Controlled Fusion*, **1**, Edition: 2, Plenum Press, New York and London (1984).

[24] A. Degasperis, D. D. Holm, and A. N. W. Hone, *A New Integrable Equation with Peakon Solutions*, Theor. Math. Phys. **133**, 2 pp. 1463–1474 (2022).

[25] D. D. Holm and A. N. W. Hone, *A Class of Equations with Peakon and Pulson Solutions*, J. Nonlin. Math. Phys. **12**, 1 pp. 380–394 (2005).

[26] P. K. Karmakar and P. Das, *Nucleus-acoustic waves: Excitation, propagation, and stability*, Phys. Plasmas, **25**, 082902 (2018).

[27] J. A. Bittencourt, *Fundamentals of Plasma Physics*, Springer Science + Business Media, **3rd** Edition, New York (2004).

[28] A. Hasegawa, *Plasma instabilities and nonlinear effects*, vol. 8, Springer-Verlag, Berlin, Heidelberg (1975).

[29] D. A. Gurnett and A. Bhattacharjee, *Introduction to plasma physics*, Ed. 2, Cambridge University Press, Cambridge, United Kingdom (2017).

[30] Y. Wang and J. Du, *The viscosity of charged particles in the weakly ionized plasma with power law distributions*, Phys. Plasmas **25**, 062309 (2018).

[31] A. A. Ahamed, S. P. Subramanian, A. M. Rahman, A. K. Raja, K. Mahalakshmi, T. K. Thirumalaisamy, *Study of solar activities with a Halo CME on 17 Feb 2023 event*, New Astronomy **114**, 102312 (2025).

[32] B. D. DePaola, *Diodes and transistors* in *practical analogue, digital, and embedded electronics for scientists*, IoP Publishing, Bristol, UK (2020).
[33] J. D. Huba, *NRL plasma formulary*, Plasma Phys. Div., Naval Research Laboratory, Washington, DC 20375 (2011).
[34] S. Dutta and P. K. Karmakar, *Sheath plasma instability in inverted fireballs*, Chaos Solit. Fract. **186**, 115259 (2024).

# APPENDIX A:
# LINEAR AND QUASILINEAR SPR FEATURES

It may be expedient and convenient for the reader to highlight some of the relevant parameters of the linear SPI analysis using various formalisms alongside investigated characteristics alongside their corresponding quasilinear counterparts.

*(a). Condition for linear SPR structure*

The purely linear perturbation (small-scale) formalism applied in this work assumes that a perturbation in a relevant physical variable of the model system is so small that its rate of change varies directly as its instantaneous magnitude. For example, if a physical parameter, $F$ undergoes a linear perturbation, $F_1$, then its spatiotemporal rate is proportional to its instantaneous state, mathematically given as

$$\frac{dF_1}{d(r,t)} \propto F_1(r,t). \tag{A1}$$

Solving this proportionality equation with a constant on the RHS yields the exponential form of perturbation variation, i.e., $F_1 = F_{10}\exp(-i(\omega t - kr))$ used in this SPR analysis within an IFB.

The above formalism allows us to examine whether the SPI pulses grow and reflect from the sheath or leak through it, by comparing the magnitude of the general solution of the DR (Eq. (32)) to the generic expression of $\omega_{spe}$ [33]. An experimental determination of the charge number density at the sheath ($n_s$) would help to figure out the magnitudes of other plasma parameters, and vice-versa. These equations can also be used to evaluate the limiting and threshold values of the relevant plasma parameters within which the IFB system can form the evanescent SPI characteristics.

The DR solution and the generic expression for $\omega_{spe}$ [33] can be arranged analytically to yield the linear SPR condition given as

$$\left(\frac{1}{m_e n_c}\right)\left(\xi k^2 + \frac{2\beta}{r^2}\right)\exp(\gamma r^2) \leq 5.64 \times 10^4 \sqrt{n_s}. \tag{A2}$$

If the LHS in Eq. (A2) is greater than the RHS at $r = R$, where $R$ is the radial distance of sheath from IFB center, the resonance condition will not be fulfilled and no SPR excitation will occur. In other situations, the SPR may still develop and reflect with LHS greater than the RHS, occurring at $r \neq R$, but it must be somewhere at $r < R$. It is noteworthy that, during the resonance, both $r$ and $k$ are interdependent through the linear SPR condition as $rk = 1/4\pi$.

In summary, the various possibilities derivable from Eq. (A2) are highlighted as follows

$$\omega_i(r = R) \leq 5.64 \times 10^4 \sqrt{n_s} \text{ (SPR occurs and gets reflected from the sheath)}, \tag{A3}$$

$$\omega_i(r = R) > 5.64 \times 10^4 \sqrt{n_s} \text{ (No SPR occurs)}, \tag{A4}$$

$$\omega_i(r < R) \leq 5.64 \times 10^4 \sqrt{n_s} \text{ (SPR occurs and gets reflected from the sheath)}, \tag{A5}$$

$$\omega_i(r < R) > 5.64 \times 10^4 \sqrt{n_s} \text{ (SPR may not occur at all)}. \tag{A6}$$

The new $\omega_i$-expression during the linear SPR behaviors in terms of $r$ (with $k = 1/4\pi r$)) as an independent variable mathematically reads as

$$\omega_i(r) = -\left(\frac{1}{m_e n_c}\right)\left(\frac{2\beta}{r^2}\right)\exp(\gamma r^2). \tag{A7}$$

Similarly, new $\omega_i$-expression during the linear SPR behaviors in terms of $k$ (with $r = 1/4\pi k$)) as an independent variable is mathematically given as

$$\omega_i(k) = -\left(\frac{1}{m_e n_c}\right)(32\pi^2 \beta k^2)\exp\left(\frac{\gamma}{16\pi^2}\left(\frac{1}{k^2}\right)\right). \tag{A8}$$

Hence, at the time of the SPI, the parameters $r$ and $k$ are interdependent. During the rest of the IFB operation, these two parameters may behave independently. Apart from other parametric values, $r$ and $k$ together yield the angular frequency ($\omega_i$) magnitude.

*(b). Condition for quasilinear SPR structure*

The linear formalism discussed before is based upon the small-amplitude perturbation assumed in the analysis. Increasing the order of perturbation, however, may break the linearity condition as the rate of change of the perturbation and its instantaneous order (magnitude) are no longer proportional. This minimum magnitude of $F_1$ for which the proportionality relation (Eq. (A1)) loses its validity yields the quasilinearity. Unlike the linear treatment there is no pre-assigned form of perturbation and the same must be derived subsequently. A plasma parameter undergoing quasilinear perturbation is written as

$$F = F_o + \epsilon F_1 + \cdots \tag{A9}$$

Where, $F_o$ is the unperturbed parametric state, $F_1$ is quasilinearly perturbed state, and $\epsilon$ is a small parameter further denoting the amplitude of the quasilinear perturbation. Besides, $\epsilon$ also denotes a balanced strength between dispersion and convection [34]. It should be noted that further increase in the order of perturbation leads to nonlinearity in the system. A table comprising of various items showing the contrast between the linear and quasilinear formalism is presented below to highlight the difference between the two.

**TABLE AI. LINEAR VS. QUASILINEAR FORMALISMS**

| S. No. | Item | Linear | Quasilinear | Remark |
|---|---|---|---|---|
| 1 | Adopted space | Wave space | Coordination space | It depends on the space of interest and relevance |
| 2 | Perturbation formalism | Fourier-type | Lowest order but moderated with $\epsilon$ | Difference is due to fluid nonlinearity |
| 3 | Order of perturbation | Lowest order (with no $\epsilon$) | Lowest order (with $\epsilon$) | $\epsilon$-moderation is in the quasilinear case |
| 4 | Outcome | Generalized linear DR | Multi-order PDE | Multi-order harmonics excite in the quasilinear case |
| 5 | Solution nature | Linear in wave space | Quasilinear in coordination space | Weak nonlinearity is in the quasilinear case |
| 6 | Superposition principle | Applicable | Not applicable | Linear transformations obey superposition |
| 7 | Fluid convection | Forbidden | Allowed | Convective dynamics is in the quasilinear case |
| 8 | Fluid dispersion | Allowed | Allowed but dictated by nonlinearity | Dispersive and nonlinear fluid nature prevails |
| 9 | Parametric interaction | Forbidden | Allowed | It ignores the parametric excitation of other modes |

| 10 | Boundary condition | System dependent | $F_1(r \to \infty) = 0$, $F_1(t \to \infty) = 0$ | Important for feasible results |
| 11 | Number of modes | Multiple | Normal acoustic mode | Each DR root is mathematically a possible mode |
| 12 | Mode saturation | Simple | Complex | Multi-order convective interactions are in the quasilinear case |